\documentstyle[epsf]{l-aa}

\begin{document}

   \thesaurus{**         
	      (**.**.*;  
	       **.**.*;  
	       **.**.*;  
	       **.**.*;  
	       **.**.*;  
	       **.**.*)} 
   \title{The formation of cD galaxies.}


   \author{A. Garijo  \inst{1}, E. Athanassoula  \inst{2} \& C.Garc\'{\i}a-G\'omez  \inst{1}.}   

   \institute{Dep. Inform\`atica. Escola T\`ecnica Superior 
	      d'Enginyeries. Univertitat Rovira i Virgili,
	      43006 Tarragona, Spain
  \and	     
  Observatoire de Marseille, F-13248 Marseille Cedex4, France}

  \offprints{C. Garc\'{\i}a G\'omez}

   \date{ Received 96 / Accepted 97}

   \maketitle

   \begin{abstract}
 
We present N-body simulations of groups of galaxies with a
number of very different initial conditions. These include spherical isotropic, 
nonspherical anisotropic
collapses and virialised spherical systems. In all cases but one 
the merging 
instability leads to the formation of a giant central galaxy in the center of 
the group. The initial conditions of the exception are such that no  
galaxies are present in the central part of the group. Thus some
central seed of material is necessary to trigger the formation of a giant
 central galaxy. We concentrate 
on the properties of these giant central galaxies. Spherical virialised 
systems give rise to relatively round and isotropic 
systems, while aspherical initial 
conditions give rise to triaxial objects with anisotropic velocity 
dispersion tensors. In the latter cases the orientation of the resulting 
central galaxy is well correlated with that of the initial cluster.
We compare the projected properties of the objects formed with the
properties of real brightest cluster member galaxies. The surface density 
profiles are in good agreement with the
observed surface brightness profiles. In the case of extended
virialised groups the projected properties of the giant central
galaxy are the same as the properties of cD galaxies. These include a halo of
luminous material and a nearly flat velocity dispersion profile.

      \keywords{ galaxies: elliptical and lenticular, cD -- galaxies: kinematics and dynamics -- galaxies: interactions -- methods: numerical.}
      
   \end{abstract}


\section{Introduction.}

The center of galaxy clusters are usually dominated by very massive 
($\sim~10^{13} M_\odot$) and extended ($\sim 300$ kpc) galaxies, 
called brightest cluster members, or D or cD galaxies, whose particular 
physical properties require a distinct formation scenario. More detailed 
information on these objects is given in 
the reviews by Tonry (1987), Kormendy \& Djorgovski (1989) and 
Schombert (1992).
Four theories have been proposed so far to explain the properties of these 
central 
dominant galaxies.  

The first theory is related to the presence of cooling flows in clusters of 
galaxies (Cowie \& Binney 1977; Fabian \& Nulsen 1977). If the central 
cluster density is high enough, intracluster gas can gradually condense and 
form stars at the bottom of the potential well. 
Andreon et al. (1992), however, show that colour gradients are small or absent, 
while McNamara \& O'Connell (1992) find colour anomalies only in the inner 
5-10\% of the cooling radii estimated by X-ray observations. Furthermore, they 
find that the amplitudes of these colour anomalies imply star formation rates
that account for at most a few percent of the material that is cooling and
accreting on the central galaxy, if the initial mass function is the same as 
that of the solar neighbourhood.

The second theory involves tidal stripping. Cluster galaxies that pass near 
the center of the cluster may be stripped by the
tidal forces arising from the cluster potential or the potential of the central
galaxy itself. The stripped  material eventually falls to the center of the 
potential 
well, where the giant
galaxy resides, and may be responsible for the halo of cD galaxies. This 
theory
was first proposed by Gallagher and Ostriker (1972) and later developed by
Richstone (1975, 1976). It can explain the halos of cD galaxies, but it is
unable to explain the differences between D galaxies, which are central dominant
galaxies with no halo, and cD galaxies. Moreover, observations show that
the velocity dispersion of the stars in cD halos is three times smaller than 
the
velocity dispersion of galaxies in the cluster, and so this theory should  
work out how tidally stripped material is slowed down as it builds up a cD 
halo.

The third theory links the formation of the central galaxy to progressive 
mergings or 
captures of less massive galaxies by the central object of a cluster. This 
theory is known 
as ``galactic cannibalism" and was first proposed by Ostriker \& Tremaine 
(1975) and later
developed by Ostriker \& Hausman (1977).  Merging might account for the
formation of the central parts of first ranked galaxies with a de
Vaucouleurs profile, since such a profile is often found in the simulations of 
galaxy mergers (Barnes \& Hernquist 1992 and references therein). 
Photometric observations (Schombert 1987) discard the analytical approach of
Ostriker and Hausman (1977), which is based on homology.
The observation of multiple nuclei in central galaxies is often cited as 
evidence in favour of the merging theory. Nevertheless, the rates of mass
increase which are obtained by analyses of samples of central galaxies with 
multiple nuclei (Lauer 1988; Merrifeld \& Kent 1991; Blakslee \& Tonry 1992) 
are more in agreement 
with a weak cannibalism than with a strong one. The observations of Thuan \& 
Romanishin (1981), Morbey \& Morris (1983) and Malumuth \& Kirshner (1985) 
also give support to the theory of galactic cannibalism.

As a fourth alternative, Merritt (1983, 1984, 1985) suggests that the essential 
properties
of cD galaxies are determined no later than cluster collapse. At later stages 
frequent 
merging between galaxies would be inhibited by the relatively high 
velocities 
between galaxies and the high fraction of the mass in a common background 
halo. Furthermore Merritt argues that truncation of galaxy halos during 
cluster collapse should make time scales for dynamical friction longer 
than a Hubble time and thus ``turn off" subsequent evolution in the cluster.

In the eighties the dynamics of clusters of galaxies were explored 
by a number of studies which use different techniques and which 
include a variety of physical phenomena via numerical recipes. Such 
approaches have been criticized by Garc\'{\i}a-G\'omez, Athanassoula \& Garijo (1996),
who have compared a few of them to fully self-consistent simulations
and have found them to be of very unequal quality.

More recently, with the advent of modern supercomputers, 
self-consistent simulations are possible. Funato et al. (1993) followed 
the evolution of 65536 particles
with the special purpose GRAPE-3 machine. Mass was distributed between galaxies
and a cluster background. The density profiles of both galaxies and background
followed Plummer distributions with different scale sizes. In this simulation 
stripping was more important than merging for the evolution of the galaxies. 
In another simulation, however, Bode et al. (1994), using 40000 particles, 
found that merging was
more important. In both works a central dominant object is formed as the 
result of the evolution of the systems. The differences in their results can be 
ascribed to the 
different density profiles and mass distributions selected to represent the 
galaxies (Bode et al. 1994). Bode et al. also looked for multiple nuclei in 
their simulations. In cases where the common halo has initially 50\% of the 
total mass in the cluster they find that multiple nuclei are seen at least 
20\% of the time, with a maximum of ~40\% at ~11 Gyrs, a number in agreement 
with what is expected from the 
projected surface density distributions. Bode et al. also showed that 
increasing the common halo mass slows the merging rate. For 90\% of the 
mass in a common halo the merging time is longer than the Hubble time.

Our simulations are fully self-consistent, but, while most simulations 
concentrate on 
the dynamics of the cluster as a whole, our aim is to study the formation of 
the central dominant galaxy. Our simulations and their initial conditions are 
presented in section 2 and their evolution is discussed in section 3. In section 
4 we study the properties of the central object, both in three dimensions and 
projected, and compare them, whenever possible, with the observed properties 
of brightest cluster members. Finally we summarise our results in section 5.

\section{Initial conditions and computational method.}

\begin{figure*}
\epsfxsize=6.5truein
\centerline{\epsffile{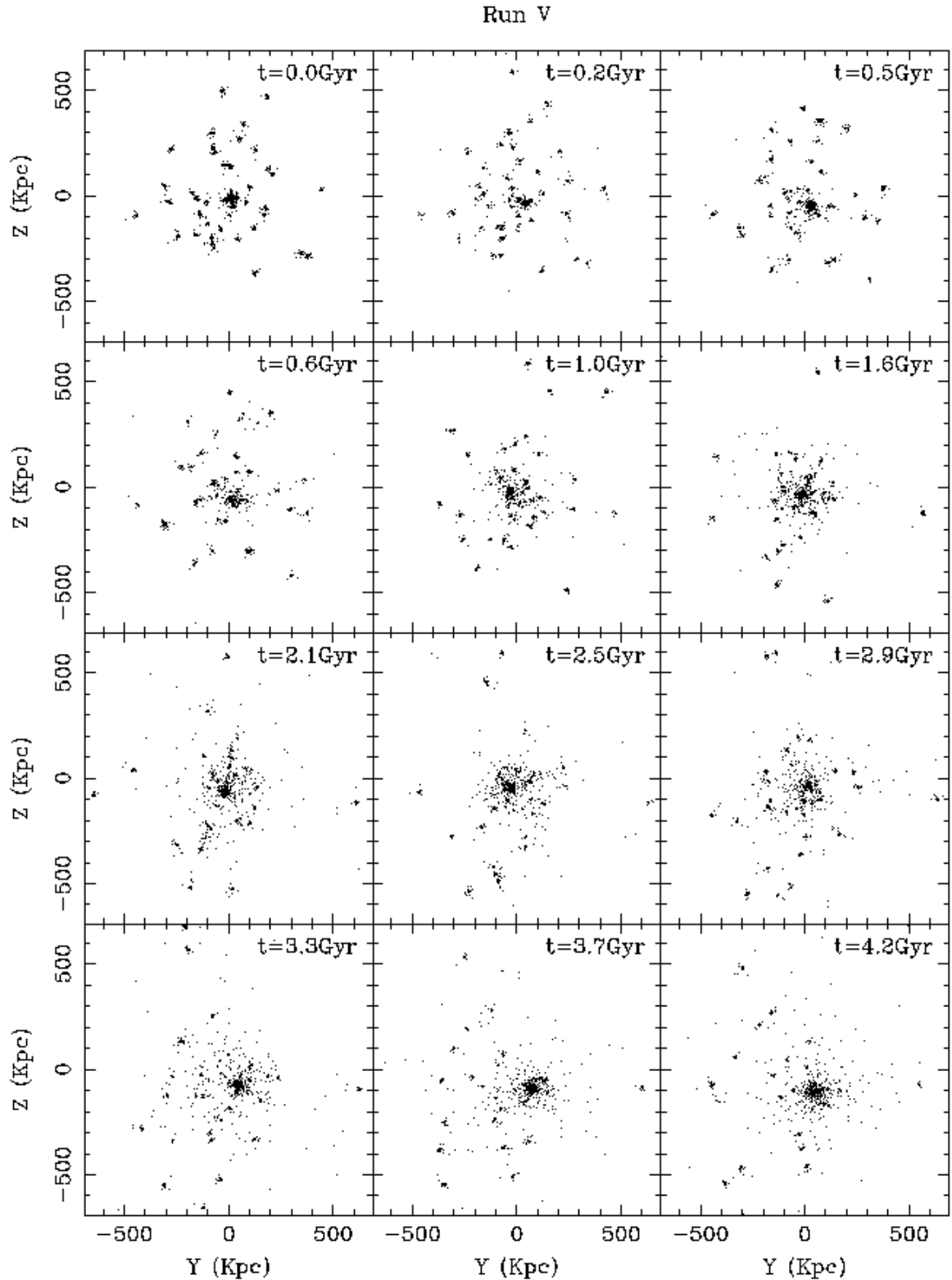}}
\caption{Time evolution of Run V. Note the rapid formation of the 
central dominant galaxy from the initial seed of galaxies and the creation 
of a uniformly distributed background from the material stripped from the
satellite galaxies by the tidal forces of the cluster. }

\end{figure*}

We have performed a series of N-body simulations with the purpose of studying 
the formation of central dominant galaxies and in particular of cD 
gala\-xi\-es. 
In all cases we used $45000$ particles representing at the onset $50$
identical galaxies of $900$ particles each. This is somewhat higher or
of the same order as the corresponding number used by Funato et
al. (1993) and by Bode et al. (1994) and, as shown by Garcia-Gomez et
al. (1996), is sufficient for our purposes. The particles in a given galaxy
were initially taken to follow a Plummer distribution of core radius $0.2$
and unit mass. For all the simulations, except for Vh, the radial
distances from the center of the group to the galaxy centers were
picked at random between 0 and $R_c$. For simulation Vh the central
part of the sphere contained no galaxy, i.e. the radial distances were
picked between 0.5$R_c$ and $R_c$. In the case of non-spherical
initial conditions
the $X$ coordinate of the distance of each galaxy from the
group center is multiplied by some appropriate constant. The essential
information on the initial conditions can be found in Table~1. 
{\sl Column}
1 gives the run identifier and {\sl Column} 2 the radius of the sphere
initially containing all the galaxies. In {\sl Column} 3 we can find the 
ratio of initial kinetic energy to the absolute value of the potential energy 
of the galaxies seen as point masses. A value of $0.0$ for this ratio stands 
for collapsing systems, while a value of $0.5$ stands for initially virialised
systems. In the latter case the velocity dispersion of the bulk motion of 
the galaxies is taken to be
isotropic. {\sl Column}~4 gives the ratio of the initial velocity 
dispersion of the galaxies within the cluster to the velocity dispersion of 
the particles within a galaxy and, finally, {\sl Column}~5 gives the axial ratios 
of the system. 

\begin{table}
\caption{Initial conditions}
\begin{flushleft}
\begin{tabular}{lllll} \hline
Name & $R_c$ & $K/\mid P \mid$ & $\sigma_c/\sigma_g$ & $x:y:z$\\ \hline
C1 & $30$ & $0.0$ & $0.0$ & $1:1:1$\\
C2 & $30$ & $0.0$ & $0.0$ & $1:1:1$ \\
Cp & $30$ & $0.0$ & $0.0$ & $2:1:1$\\
Co & $30$ & $0.0$ & $0.0$ & $0.5:1:1$ \\
V & $20$ & $0.5$ & $1.4$ & $1:1:1$\\
Vh & $20$ & $0.5$ & $1.0$ & $1:1:1$\\
Vc1 & $10$ & $0.5$ & $1.9$ & $1:1:1$\\
Vc2 & $10$ & $0.5$ & $1.8$ & $1:1:1$\\\hline
\end{tabular}
\end{flushleft}
\end{table}

Runs C1, C2, Cp and Co (``p" for prolate and ``o" for oblate) are initially collapsing 
systems, 
while runs V, Vh, Vc1 and Vc2 (``h" for hollow and ``c" for compact) are
initially virialised systems. We shall, for brevity, often refer to the 
former simply as ``collapsing", rather than ``initially collapsing", and 
to the latter simply as ``virialised". 
Runs~V and Vh have similar initial global
conditions but, for reasons which will be discussed later, run Vh was
initially depleted of galaxies in the central part. Since the number of 
galaxies in all the runs is the same, this means that there are more galaxies 
in the 
outer parts of the group in run~Vh than in run~V. Runs~Vc1 and~Vc2
are also virialised systems, but in these cases the initial radius of the
sphere containing the galaxies is half that of run V.
In all these initial conditions all the mass is initially bound to
galaxies. Simulations where a fraction of the mass is in a common halo will
be discussed in a future paper. 

For the
collapsing simulations C1 and C2 and the virialised ones Vc1 and Vc2 we used 
the same global initial conditions, but
different initial seeds, in order to check for a possible influence of 
the realisations
on the final results. The collapsing systems of run~Cp 
and run Co 
were initially anisotropic systems and were performed with the aim of studying 
the possible influence of the initial shape of the system on the final 
properties of the central galaxy. Run~Cp is an initially prolate system where
the initial size of the $X$ axis is doubled, while run Co is an initially oblate
system for which the initial size of the $X$ axis is halved. 

We followed the evolution of these groups using a version of the Barnes and
Hut treecode (Barnes \& Hut 1986), particularly adapted for a Cray computer
(Hernquist 1988). The time step was taken to be equal to $0.0075$ and
the softening length equal to $0.05$, which is of the order of the
mean interparticle distance in the initial galaxy. This ensured an
energy conservation
better than $10^{-3}$. Each simulation was continued for $4000$ steps, i.e. a 
total time in simulation units of $30$. One simulation lasted about
150 hours on a CRAY 2L. In this paper we will use the
computer units $M_{gal}  = R_{gal} = G = 1$, where $M_{gal}$ is the initial 
mass in each 
galaxy, $R_{gal}$ is their initial radius and $G$ is the gravitational
constant.
To compare with the observations these can be converted to real units by
assigning a mass and a radius to each galaxy. In the following we will assume 
$M_{gal} = 5 \times 10^{11} \,\, M_\odot$ and $R_{gal} = 30\,\,$ kpc, which 
gives $1.40 \times 10^8 \,\,$ years and $208\,\,\hbox{kms}^{-1}$ for the 
units of time and velocity respectively. It is obvious that this choice, 
albeit reasonable, is not unique, and that other neighbouring values would 
have been equally well acceptable. This should be kept in mind when comparing 
with observational data, hence agreements to within a factor of two should be 
considered quite satisfactory.

\section{Evolution of the simulations.}

We show in Fig.~1 the time evolution of run V. A central
galaxy is quickly formed beginning with the galaxies initially near the
center of the group. Later this galaxy grows from the rest of the members of
the cluster, often literally swallowing up an entire companion, as
suggested in the galactic cannibalism picture (Ostriker \& Hausman 1977).
Galaxies also lose material due to tidal forces. This mass is accreted to
the center of the potential well which is occupied by the central giant
galaxy. The evolution of the other simulations is similar to that of 
run~V in the sense that a central galaxy is quickly formed, growing, 
however, at different rates in the different simulations. The sole exception is 
run Vh, where no central object is formed.

In Fig.~2 we plot the mean distance of the galaxies from the center of the 
cluster, $R_g$, as a function of time for all our simulations. The central galaxy 
is not taken 
into account in this average, and we also do not introduce a weighting 
by mass. The size of the 
group diminishes steadily
in all the collapsing systems but at different rates. Runs C1 and C2, which are 
two different realisations of the same initial global conditions, evolve at 
similar rates until the time of maximum collapse is reached at $t \sim 2.8$ Gyr. 

\begin{figure}

\epsfxsize=3.5truein
\centerline{\epsffile{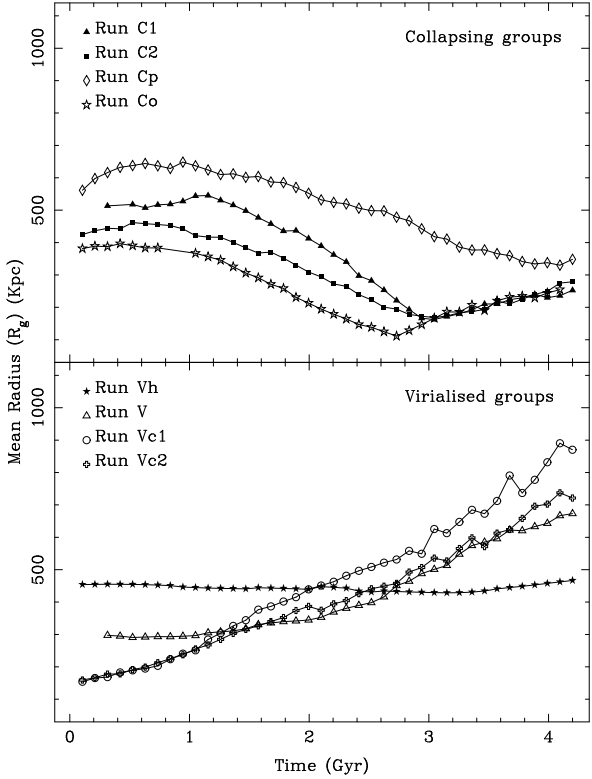}}

\caption{Time evolution of the mean radius of the group of surviving
galaxies for each of our simulations. In the top panel we show the evolution
of this parameter for the collapsing clusters and in the lower panel the
evolution for the initially virialised clusters. Symbols for collapsing groups: run~C1 filled triangles, run~C2 filled squares, run~Cp diamonds, run~Co stars. Virialised groups: run~Vh filled stars, run~V triangles, run~Vc1 circles, run~Vc2 swiss crosses. }

\end{figure}

After this 
time both systems suffer the same small expansion. The similarity between the 
results of runs C1 and C2 indicate that the statistical fluctuations in 
the initial conditions have a small influence on the later evolution of the 
group. This same conclusion 
can also be reached by comparing runs Vc1 and Vc2, for which the differences are even
smaller. The system of run Co is maximally collapsed at a slightly 
earlier time. Because its extension along the $X$ axis is smaller than that 
of the spherical symmetric systems by a factor of 
$2$, it 
starts out smaller and denser, and hence it collapses faster. In the 
case of run~Cp on the other hand, the extension along  
the $X$ axis is greater than that 
of the spherically symmetric systems by a factor of 2, so the system is less dense 
and the collapse rate is 
slower. For all virialised groups, except for run Vh, the 
mean radius of the group increases steadily. As there are no single galaxies 
left in the central parts 
the mean radius of the distribution gets bigger. Moreover, a few galaxies 
acquire enough speed to reach large distances from the system. These factors 
increase the 
mean radius of the system. This is not the case for run Vh, where no central 
dominant galaxy is formed, and for which the mean radius of the cluster 
stays roughly constant. 

Fig.~3 shows that the total number of galaxies diminishes steadily as a 
function of time, but at different rates for each simulation. 
Again runs C1 and C2, as well as runs Vc1 and Vc2, behave in a very similar 
way. For run~Cp, initially a bigger prolate 
system, galaxies need longer times to reach the center and so the number of 
mergings is lower. On the other hand the
number of mergings for run Co is similar to the number of 
mergings in the spherically collapsing systems, i.e. those of runs~C1  and~C2.
This result is unexpected and we have no clear explanation to offer. 
For the virialised groups the two extreme cases correspond to
run Vh, where only seven galaxies disappear, and runs Vc1 and Vc2, where this number 
decreases very rapidly at the start and levels off towards the end of the 
simulation. The difference between run~V on the one hand and runs Vc1 and Vc2 on 
the other  is due to the fact that the galaxies in runs Vc1 and Vc2 are 
closer together, an effect which seems to be stronger than the relatively 
larger relative velocities 
between galaxies. Note also the strong difference between run~V 
and run~Vh, which are simulations with similar global properties, except for the 
fact that the latter simulation has no galaxies in its central regions.
This shows that the presence of an initial central seed of galaxies initiates
the merging instabilities and the formation of the central object via mergings. 
A central giant galaxy is formed in only three of our four cases 
with similar virialised initial conditions. For the case of Run~Vh the seven 
galaxies that have disappeared by the end of the simulation are not in a 
central object. A background is formed by diffuse material that has been 
stripped from the galaxies, while some galaxies merge in the external parts. 
On the contrary in run~V, where all the evolution of the system is 
driven by the central object, no mergings between satellite galaxies occur.

\begin{figure}

\epsfxsize=3.5truein
\centerline{\epsffile{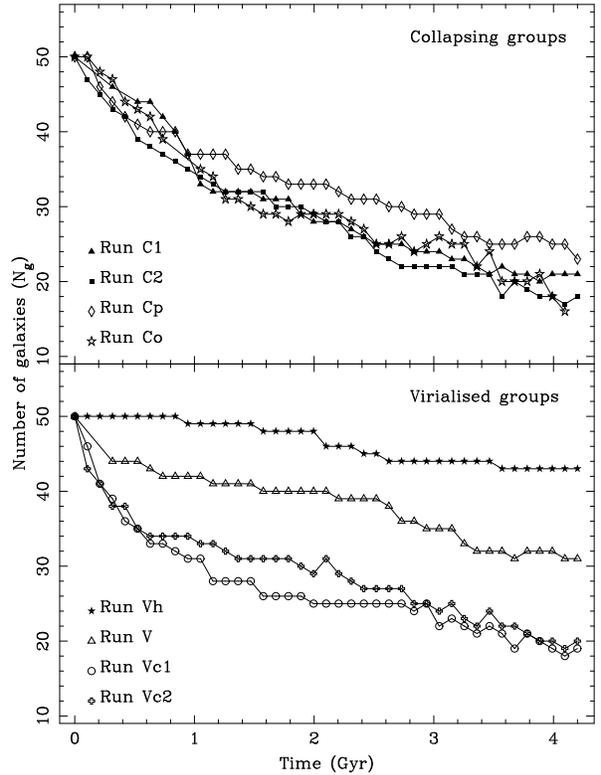}}

\caption{Time evolution of the number of surviving galaxies in the group
for each of our simulations. Panels are as in Fig.~2. Note the rapid decrease 
of this number for runs Vc1 and Vc2. Symbols as in Fig.~2. }

\end{figure}

In Fig.~4 we show the time evolution of the mass of the central object. This 
mass grows steadily in all 
simulations. For the collapsing 
systems the central object grows rapidly, at a rate which is in most cases 
roughly constant with time. Furthermore, the mean rate is higher than that of 
the virialised groups of the same initial size (runs V and Vh). For the 
virialised systems Vc1 and Vc2 
the rate of the evolution is very rapid at the start of the 
simulations, due to their small initial radius. Galaxies last 
longer in the virialised groups of run~Vh and run~V. Actually in run~Vh there 
is no central object and the mass quoted in this figure corresponds to a 
distributed background. In run~V secondary galaxies also lose material via tidal 
stripping, but there is a central object growing by accretion of this material 
and by merging of some satellite galaxies. Differences in the realisations 
of the initial conditions affect somewhat the result for the collapsing 
simulations C1 and C2, but hardly so for the virialised ones Vc1 and Vc2.
\begin{figure}

\epsfxsize=3.5truein
\centerline{\epsffile{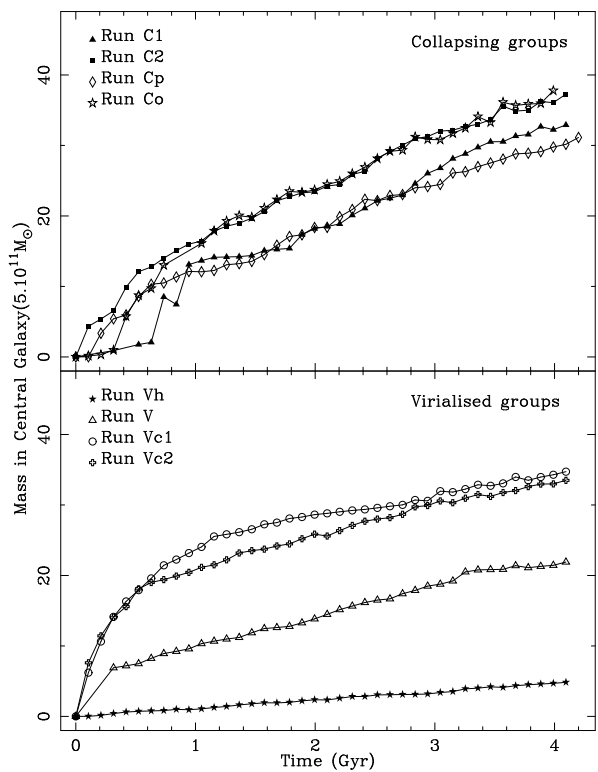}}

\caption{ Time evolution of the mass of the central object. This mass
always increases steadily with time, except for Run Vh, where no central galaxy
is formed and the mass quoted corresponds to the mass of the distributed
background. Symbols as in Fig.~2.
}

\end{figure}

In order to differentiate between growth of the central galaxy by mergings and 
growth by accretion of material from other galaxies and to measure the amount 
of stripping in each simulation, we plot in Fig.~5 the time evolution of the 
parameter $\Delta m$ defined as:
$$ \Delta m  = \frac{M_{0g}-M_g} {M_c}$$
where $M_g$ is the mass in galaxies excluding the central object, $M_{0g}$ is 
the mass that this same number of galaxies should have if there were no 
stripping or merging between them (i.e. the mass of the same number of galaxies 
in the beginning of the simulation) and $M_c$ is the mass of the central object. 
If there are some mergings between the secondary galaxies then $M_g$
will be greater than $M_{0g}$ and this parameter will take on negative values. 
If the dominant effect is the stripping by the tidal
potential of the group and merging only takes place between the secondary 
galaxies and the central object, $ \Delta m$ will take positive values. The 
greater the value of this parameter, the greater the amount of mass loss by 
stripping in the secondary galaxies.
Fig.~5 shows that the evolution of $ \Delta m$ depends strongly on the initial 
conditions of the system and, for the case of the collapsing simulations, 
even on the realisation of the initial conditions. In some collapsing groups 
some mergings between 
secondary galaxies take place, and these dominate the evolution of the group 
until there is a big central object formed in the center. Thereafter we  
only find mergings with this giant galaxy and stripping of the secondary 
galaxies. In the virialised systems there is no merging between the secondary 
galaxies. These spiral to the center, losing some material, and finally merge 
with the central object. We may also note that the effect of stripping is more 
pronounced in run~V, which is the extended virialised system. The core radius 
of the galaxies, defined as the 
radius containing $35\%$ of the most bound particles, remains nearly constant 
during the simulation, with variations of less than $0.05\%$. The velocity 
dispersion of this set of 
particles, which can be taken as a measure of the central velocity dispersion 
of the galaxy, suffers a small decrease ($0.15\%$). 
\begin{figure}

\epsfxsize=3.5truein
\centerline{\epsffile{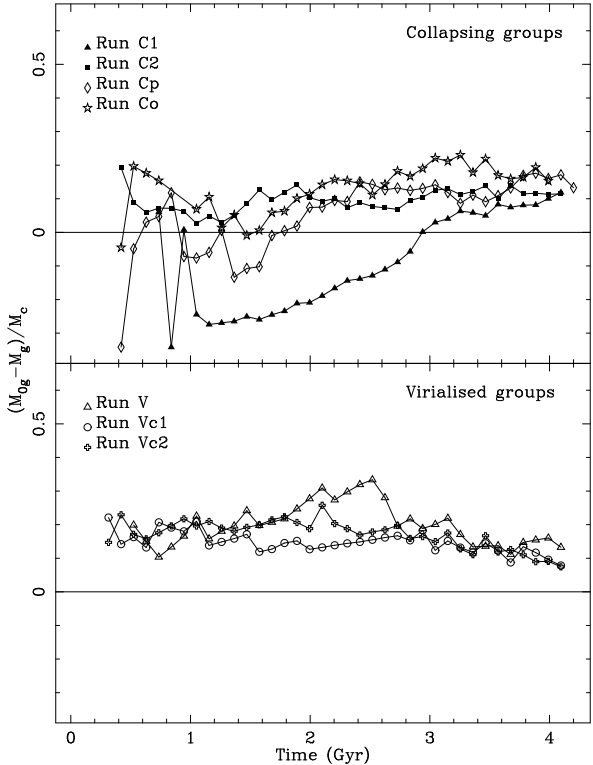}}

\caption{ Time evolution of the parameter $\Delta m$, defined in the text,
which measures the degree of evolution in the surviving galaxies. Positive
values of this parameter indicate that the galaxies are suffering stripping
of their outer parts, while negative values indicate that some merging between
the secondary galaxies is taking place. Symbols as in Fig.~2.}

\end{figure}

In Fig.~6 we plot the time evolution of the total 
mass of the 
central galaxy and the mass increase due to stripping from the secondary galaxies. The difference between the gives
the contribution from
merging of the secondary galaxies with the central object. In order to 
obtain smoother curves, which are necessary particularly for the derivatives 
discussed in
the next paragraph, we have used $5$-point sliding means of the data.
As can be seen, 
merging is the dominant process for most collapsing cases, while for the 
extended 
virialised system (run~V) the contribution from stripping is dominant. 
It is interesting to 
note the difference between the evolutions of the central 
object in runs~C1 and~C2, which are different realisations of the same 
global initial conditions. Their evolution as a group is quite similar, but
the evolution of the central object differs considerably. While in 
run~C1 there are nearly no mergings in the first timesteps, in run~C2 the 
central object grows from the start of the simulation. This is simply due to 
the presence or absence of a couple of galaxies in the central parts. 
Furthermore, the contribution of merging and stripping to the mass of the 
central galaxy is very different in the two cases. In run C1 the contribution 
of stripped material is very small, while it is comparable to that of mergings 
in run~C2. The central galaxy 
formed in run~Co, which is the denser system, also grows quickly,
while the galaxy 
in run~Cp, the less dense group, grows slower. For the virialised 
simulations the galaxy formed in run~Vc1, the most tightly bound case,
grows very 
quickly at the start, but after some time, when nearly half of the galaxies 
have disappeared, the rate levels off. This is not true for the galaxy
formed in 
run~V, since the 
system is less dense, the tidal forces are not as strong and the secondary 
galaxies in the central parts do not lose their identity as
easily. The results for run~Vc2 are very similar to those for run~Vc1,
and thus have not been plotted.

\begin{figure}

\epsfxsize=3.5truein
\centerline{\epsffile{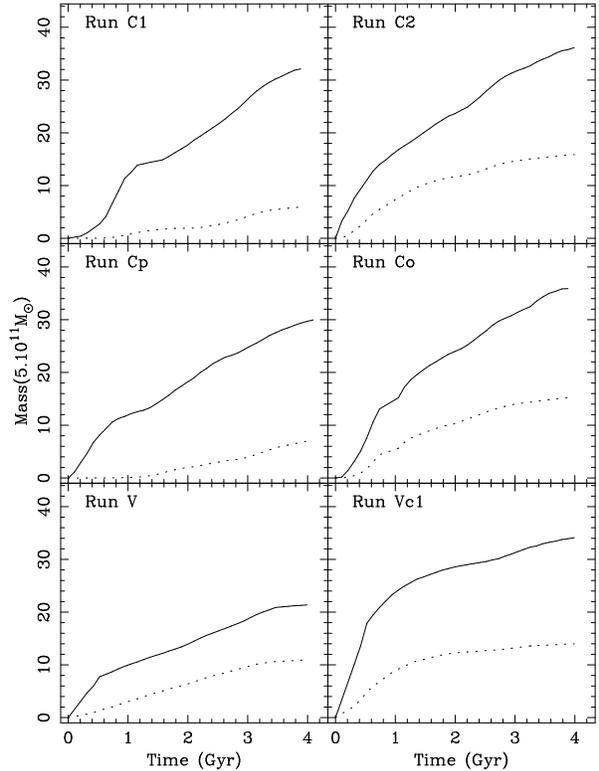}}

\caption{ The total mass in the central object as a function of time
(solid line) and the contribution to this mass from stripped material 
(dotted line). The contribution of 
stripped material is quite important in all the cases especially in the
initially virialised groups.}

\end{figure}

Fig.~7 shows the rate at which this mass increase
proceeds. The values plotted are the values obtained from those in 
Fig.~6 using a centered three point approximation for the derivatives. 
The peaks in the solid lines 
correspond to recent mergings and the peaks in the dotted lines are associated 
with massive accretion of stripped material.  In collapsing 
systems the contribution
from merging dominates over the contribution from stripping in a fair
fraction of the time. On the other hand stripping is more important
in the case of initially virialised systems. In the case of run Vc1 the
stripping is very important in the initial stages of the simulation. 
Later on, when the central object is bigger, it is the merging that 
dominates.
The most interesting case is run V, where stripping dominates during
nearly all the simulation.

\begin{figure}

\epsfxsize=3.5truein
\centerline{\epsffile{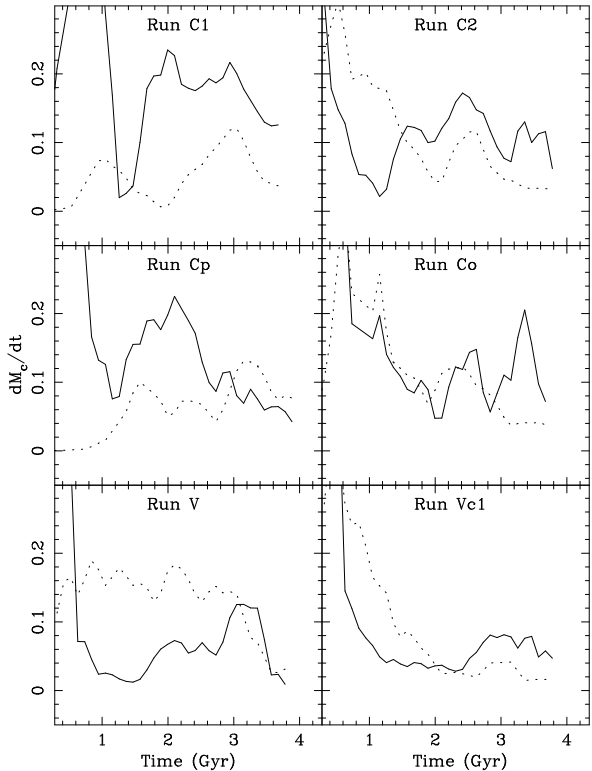}}

\caption{ Comparison of the rate of increase of mass of the central 
object 
due to merging (solid line) and stripping (dotted line). In collapsing 
systems the contribution
from merging dominates over the contribution from stripping in a fair
fraction of the time. On the other hand for the
central object formed in run V (extended virialised system) the mass 
increase is mainly dominated by stripped
material.
 }

\end{figure}

\section{The central galaxy}

\subsection{Definition of the central galaxy}

We first need to define the central galaxy in an unambiguous way. The 
information about the particles in
each galaxy is stored in a sequential way. Thus the first $900$ particles 
always correspond to the 
particles initially bound to the first galaxy, the second $900$ particles to
those initially bound to the second galaxy and so on. The process used 
to decide which particles are bound to the central galaxy at a given time is
as follows: For each galaxy, we take the initially bound $900$ particles 
and calculate the binding 
energy of each particle with respect to this subsystem. We discard all 
particles with positive energy and we repeat the process until we get a 
stable number of particles. Usually only two iterations are needed. 
We thus define the secondary galaxies at a given time step of the simulation. 
The discarded particles are not immediately incorporated into the background. 
As a second step we consider the possibility of some mass transfer between
galaxies. We
check the possibility that some of the particles that have escaped their 
initial parent galaxy are now bound to another one of the
galaxies. For each galaxy we calculate the energies
of all the particles inside a sphere of radius $R=1$ centered on the galaxy.
All the particles which are not part of another galaxy and which have 
negative energies 
are ascribed to this galaxy. If
after these two steps we find a galaxy with less than $10 \%$ of the number of
points it had initially, we do not consider it as a single entity and 
its particles are assigned to the background. After 
these two steps we are left with two sets of particles. The first set consists 
of the particles bound to some of the galaxies and the second is the set of 
particles not bound to any galaxy. We consider this second set of particles 
separately, and, in order to distinguish which, amongst these particles, 
constitute the central object and which the diffuse background, we
calculate the energy of these particles relative to
this subsystem. Only particles with negative binding energy are considered as 
the particles forming the central object. 

Finally we also take into account the possibility of merging between the 
satellite galaxies and the central galaxy, or between two galaxies. We check 
if a galaxy
of core radius $r_{c1}$ and central velocity dispersion $\sigma_1$ has
merged with a second galaxy with parameters $r_{c2}$ and $\sigma_2$ placed
at a distance $\Delta r$ and moving at a relative velocity $\Delta v$ to the
first galaxy. If both conditions 
\begin{eqnarray}
\Delta r & < & A(r_{c1} + r_{c2}) \\
\Delta v & < & B(\sigma_1 + \sigma_2)
\end{eqnarray}
are satisfied simultaneously, 
the two galaxies are merged and all the particles of the smaller galaxy are
ascribed to the bigger one, conserving at the same time the momentum of the 
binary system. After some tests the parameters $A$ and $B$ were given
the values 
$1.4$ and $0.6$ respectively.

The core radius of a given galaxy is defined as follows: We first sort the 
particles in a galaxy as a function of their binding energy and consider the 
$35\%$ which are most bound. The core radius is then defined as the smallest 
radius containing these particles. The value of $35\%$ has been obtained by 
imposing that the core radius thus obtained for the initial galaxies coincides 
with their initial core radius, i.e. 0.2. With these particles we calculate 
the mean position and velocity of the galaxy. Its central velocity dispersion 
is 
defined as the dispersion of the particles within the core radius. For the 
case of the central object, we used only the $10 \%$ most bound
particles, instead of $35 \%$ as in the satellite galaxies, in 
order to avoid an excessive number of mergings between the satellite
galaxies and the central object.

\subsection{Three dimensional properties}

The central object is divided into three concentric shells as follows: We 
rank particles in order of decreasing binding energy. We neglect particles 
in the top 5th percentile, i.e. the most bound particles, because they could be 
affected by the softening length of our simulations. In the first shell 
we include particles with binding energy between the 5th and 30th 
percentiles; in the second shell particles with binding energy between 
the 30th and 60th percentiles and in the last shell particles with binding 
energy between the 60th and 90th percentiles. We exclude the last 10\% of 
particles because these could be recently accreted material, 
and thus not in equilibrium with the rest of the galaxy. Using
only three shells we have enough particles in each
one. In this we follow the method used by Barnes (1992) in his study of the 
merger
remnants of the collision between two spiral galaxies, with the difference that 
he used only the
$75\%$ most bound particles, while we include in our analysis the $90\%$ 
most bound particles, since we are interested mainly in the 
external parts. 

\subsubsection{Shape and orientation }

Porter, Schneider \& Hoessel (1991) showed from isophotometry of 175
 brightest cluster ellipticals that in most cases their ellipticities 
increase with increasing radius. Ryden, Lauer \& Postman (1993) derived 
mean isophotal axis ratios for 119 brightest cluster ellipticals in Abell 
clusters and found from best fitting models  that their most probable 
shape parameters are $b/a=0.8$ and $c/a=0.76$. Furthermore an increasing 
wealth of evidence shows that the orientation of the brightest cluster 
ellipticals is not random, but correlates well with that of the cluster 
in which they are found (Sastry 1968, Rood \& Sastry 1972, Austin and 
Peach 1974, Carter \& Metcalfe 1980, Bingelli 1982, Struble \& Peebles 
1985, Rhee \& Katgert 1987, Lambas et al. 1988)

In order to compare the orientation and the shape of our simulations with that 
of brightest cluster members we calculate
for each of the three shells discussed in the beginning of this section the 
normalised inertia tensor defined as:
$$ {\bf D} \equiv \sum_{i=1} m_i \frac{{\bf x}_i \otimes {\bf x}_i }
{|{\bf x}_i|^2}$$
This normalised tensor gives similar results as the non-normalised one, 
while avoiding some problems with the more distant particles (Barnes 1992). 
On the other 
hand, this tensor could have problems with the particles that are near the 
center, but these particles are among the $5\%$  
most bound particles that are discarded from our analysis.

This tensor can be diagonalised to obtain the eigenvalues $(Q_1 \leq Q_2 \leq
Q_3)$ and their associated eigenvectors. The axial ratios of each shell are 
defined as follows:
\begin{eqnarray*}
\frac{b}{a}  & = & \sqrt{\frac{Q_2}{Q_3}} \\
\frac{c}{a} & = & \sqrt{\frac{Q_1}{Q_3}}
\end{eqnarray*}
\noindent
We do this for each shell and for each snapshot. In this way we are able to 
study the shape of the different shells, their relative alignment and their 
time evolution.

In Fig.~8 we show the results of the time evolution of the axial ratios 
for the three concentric shells of the 
central objects obtained in the simulation of collapsing systems. 
In the left panels we show the $b/a$ ratios and in the right panels
the $c/a$ ratios. All objects are oblate or mildly triaxial, the 
triaxiality being most evident in the 
external shell, in good agreement with the observational results of 
Mackie, Visvanathan \& Carter (1990) and Porter, Scneider \& Hoessel 
(1991). The most triaxial system is obtained in the case of 
the oblate initial conditions of run~Co. In this 
case there is also a clear difference between the ellipticity of the three 
shells, the inner shell being the roundest. There are some differences in the 
evolution of the shape parameters for the central objects formed in runs~C1 and C2. 
The inner parts of the object of run~C1 get rounder with time, while this is not 
the case for the object formed in run~C2. This might be explained by the 
different merging histories of the two simulations. 
\begin{figure}

\epsfxsize=3.5truein
\centerline{\epsffile{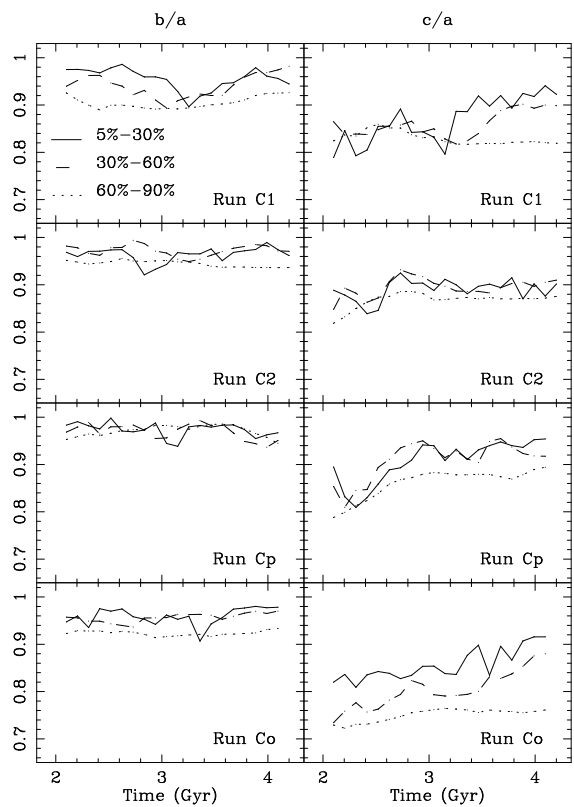}}

\caption{ Time evolution of the axial ratios of the central galaxy in
our collapsing simulations. The solid line indicates the evolution of these
ratios for the shell containing the particles with binding energy between the 5th and 30th percentiles. The dot-dashed line corresponds to the intermediate shell 
containing particles with binding energies between the top 30th and 60th percentiles, and the dotted line corresponds to the outermost shell, containing particles with binding energies between the 60th and 90th percentile. }

\end{figure}

In Fig.~9 we can see the time evolution of the axial ratios of the 
central objects formed in the initially 
virialised systems. Note that these spherically virialised systems form 
rounder objects
than those formed by spherically collapsing simulations. In the latter
cases, the galaxies in the 
collapse follow mainly radial orbits and enter the central galaxy in some 
particular direction thus affecting the shape of the central object. On the 
other hand, in the virialised systems an 
important fraction of the mass of the central galaxy comes from material 
stripped from the secondary galaxies. This material accumulates onto 
the central object at a slower rate 
and, coming from any direction, gives rise to these rounder 
objects. The differences between runs Vc1 and Vc2, which are different 
realisations of the same initial conditions, are rather small.

\begin{figure}

\epsfxsize=3.5truein
\centerline{\epsffile{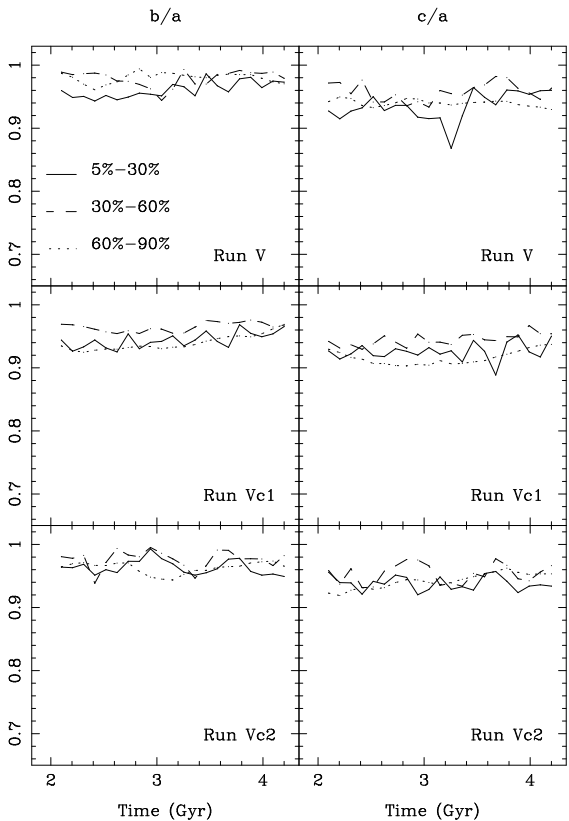}}

\caption{ Same as Fig. 8 but for the virialised simulations. In this
case all the resulting objects are nearly spherical. Lines as in Fig.~8. }

\end{figure}

Observations indicate that the brightest cluster galaxies 
seem to be more triaxial than what our simulations show,
and that their projected axial ratios have mean values lower than the values 
obtained in our simulations (Ryden et al. 1993). This may indicate that the
initial conditions for the formation of clusters of galaxies were far from 
spherical and/or very anisotropic. 

Does the orientation of the central object reflect the orientation of the 
cluster from which it initially formed? Two of our simulations, Cp and Co, 
have non-spherical initial conditions. As we already saw they form 
non-spherical 
central objects. Fig. 10 shows the angle between the minor axis of the 
initial configuration and that of each shell of the central object of run~Co 
as a function of time. We see that for all three bins there is a very 
good alignment of the central object with the initial cluster. The central 
object formed in run~Cp is much more spherical.  For that reason, in Fig. 11
we have plotted the angle between the major axis of the central 
object and the major axis of the initial configuration, as well as the angle 
between the median axis of the central object and the major axis of the 
initial configuration. Again the three shells of the central object are 
treated separately. The two major axes coincide well at 
all times for the outer shell. They correspond well most of the time for the 
median shell, and 
for some times for the innermost one. For these two shells and for the times 
when the two major axes do not coincide, it is the intermediate axis that 
corresponds to the direction of the initial major axis, especially in the inner,
more spherical regions. Similar results have been found by Rhee \& Roos (1990) in their simulations of collapsing 
small clumps of galaxies, although they find that the orientation is better
preserved in initially prolate, rather than initially oblate systems.
\begin{figure}

\epsfxsize=3.5truein
\centerline{\epsffile{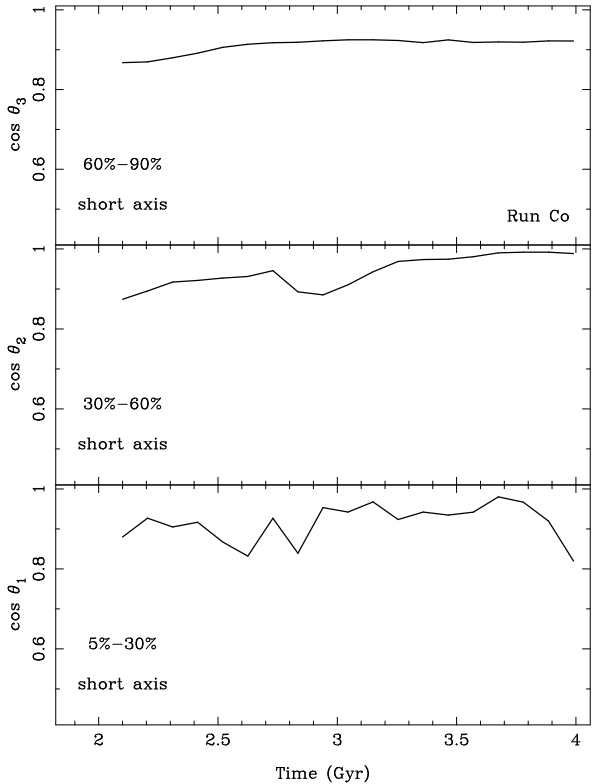}}

\caption{ Time evolution of the angle between the minor axis of the 
initial configuration and that of the central galaxy for Run~Co. In
 the upper panel the angle refers 
to the shell containing the particles with binding energy between the 5th and 
30th percentiles. The middle panel corresponds to the intermediate shell 
containing particles with binding energies between the top 30th and 60th 
percentiles, and the lower panel corresponds to the outermost shell, 
containing particles with binding energies between the 60th and 90th 
percentile.
}

\end{figure}
\begin{figure}

\epsfxsize=3.5truein
\centerline{\epsffile{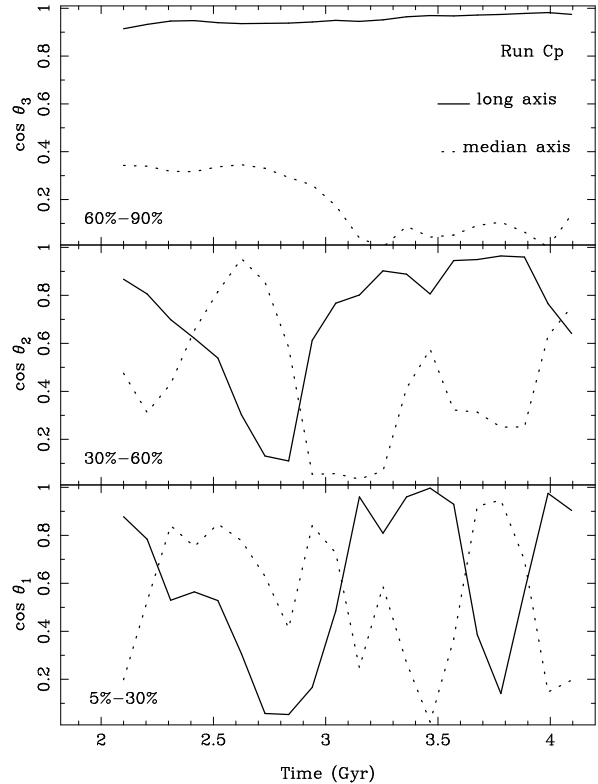}}

\caption{Time evolution of the angle between the major and median 
axes of central galaxy with the major axis of the initial configuration
for Run~Cp. The solid line corresponds to the major axis and the
dotted line corresponds to the median axis. }

\end{figure}

\subsubsection{Volume density }
 
In order to study the properties of the central galaxy as a whole we define
the principal directions of the whole galaxy as the principal axes of the 
inertia tensor of particles with binding energy between the 5th and 60th 
percentiles. The last shell is discarded because the outermost particles
often have a substantial asymmetry. The eigenvalues of this system were used to define the ellipticity of the 
galaxy as a whole and we take its principal directions as the directions of the 
eigenvectors. Using these values we fitted the three dimensional density profile
to a Hernquist profile (Hernquist 1990) using the expression given by Dubinski 
\& Carlberg (1991):
$$ \rho(q) = \frac{M_T}{2\pi} \frac{1}{c_1c_2} \frac{q_s}{q}
\frac{1}{(q+q_s)^3}$$
\noindent
where $M_T$ is the total mass, $q_s$ is a scale length,
related to the half mass ellipsoidal surface
$$q_{1/2} = (1+\sqrt{2})q_s,$$
\noindent
$q$ is the ellipsoidal coordinate
$$ q^2 = x^2 + \frac{y^2}{c_1^2}+\frac{z^2}{c_2^2}$$
\noindent
and $c_1$ and $c_2$ are the axial ratios of the whole galaxy.

The particles in the central object were sorted according to their ellipsoidal
coordinate and binned in shells, each containing $200$ particles. A good fit to the 
Hernquist profile indicates that the mass distribution is stratified
on similar ellipsoids at all radii. This profile was fitted for all the central
objects in each snapshot of the simulations. The results are shown in Fig.~12
for the collapsing groups. All the galaxies formed in the 
collapsing simulations are well fitted by the Hernquist profile. In Fig.~13 we 
show the same plot for the initially 
virialised systems. We can see that the objects formed in the strongly
bound systems (runs Vc1 and Vc2) are also well fitted by the Hernquist profile. 
Run~V is the most interesting case. The central object 
formed in this simulation is not well fitted by the Hernquist profile, thus 
indicating that the object formed under these initial conditions has a 
different structure. 

\begin{figure}

\epsfxsize=3.5truein
\centerline{\epsffile{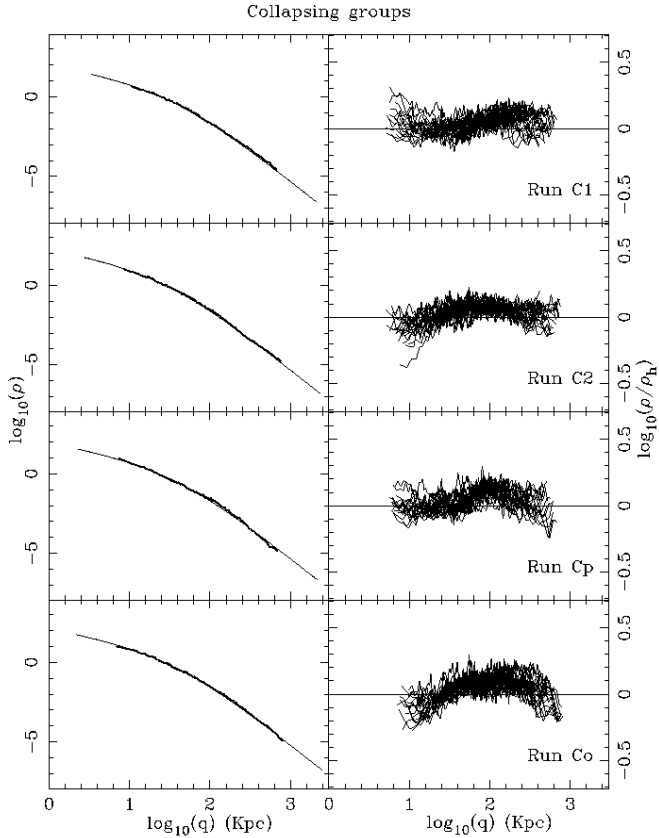}}

\caption{ Fits of the three dimensional density distribution of the
central objects formed in the collapsing simulations. In the left panel we
show fits of Hernquist profiles to the data corresponding to the last 
integration step. In the right plane we show
the deviations of the real density from the fitting function for all the radii
and all the snapshots. The small values of these deviations indicate good fits 
by this law for the entire system and thus that these central objects are 
homologous.
 }

\end{figure}
\begin{figure}

\epsfxsize=3.5truein
\centerline{\epsffile{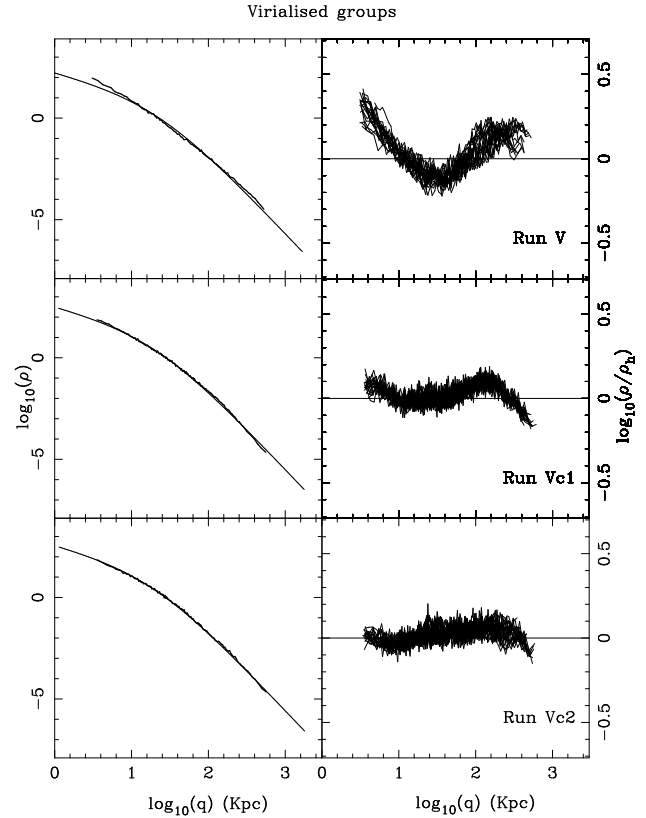}}

\caption{  Same as Fig. 12, but for the virialised simulations. Note
that the central object formed in run V is not well fitted by the Hernquist
profile, indicating a non-homologous nature for this object.
}

\end{figure}

\subsubsection{Velocity dispersion and anisotropy }

To measure the degree of isotropy in the central galaxies we use the mean 
velocity dispersions in each of the principal directions. For a spherically
virialised central object we expect similar values along each of the 
principal 
axes. In Fig.~14 and 15 we plot the results for the collapsing groups and 
virialised groups respectively. For the collapsing groups we  
see that the values in any direction vary in a very irregular way, while 
for the initially virialised simulations they remain nearly constant 
during the time span of the simulation. These differences are due to the
different evolutionary histories of the two classes of systems, as 
discussed in section~3 and shown in Fig.~4. In the collapsing groups the 
mergings occur at a roughly constant rate all through the evolution, and at
all times there is material that has not settled yet to some equilibrium. On the
other hand, for the virialised systems the central
objects are to a large extent the result of mergings during the initial 
stages of the evolution, between 
the galaxies forming the central seed. Thus the material has had more time 
to settle to equilibrium.
The addition of new material, both by merging 
and in the form of stripped material, comes at a slower rate, presumably 
slow enough so as not to alter the existing equilibrium in any crucial way. 
The presence or absence 
of irregularities is not the only difference between collapsing and 
virialised cases. For collapsing groups 
the velocity 
dispersion along the $X$ axis is systematically higher than the velocity 
dispersion along 
any other direction. This is in agreement with the fact that these are
non-spherical 
systems supported by anisotropic velocity dispersion tensors. Note also that 
the radial motions dominate over the tangential ones in all cases. This is 
due to the particular process of formation of these objects, whereby 
merging galaxies 
enter the central object following mainly radial orbits. On the other hand 
for initially virialised 
cases the velocity dispersion along the $X$ axis does not dominate over the
velocity dispersion along the rest of the principal directions, in agreement
with the fact that these objects are less ellipsoidal. The three components of 
the velocity dispersion in spherical coordinates are also nearly equal during 
all the simulations, indicating that they are isotropic systems in equilibrium 
and that there is no ordered motion of the particles which 
constitute these central objects.
\begin{figure}

\epsfxsize=3.5truein
\centerline{\epsffile{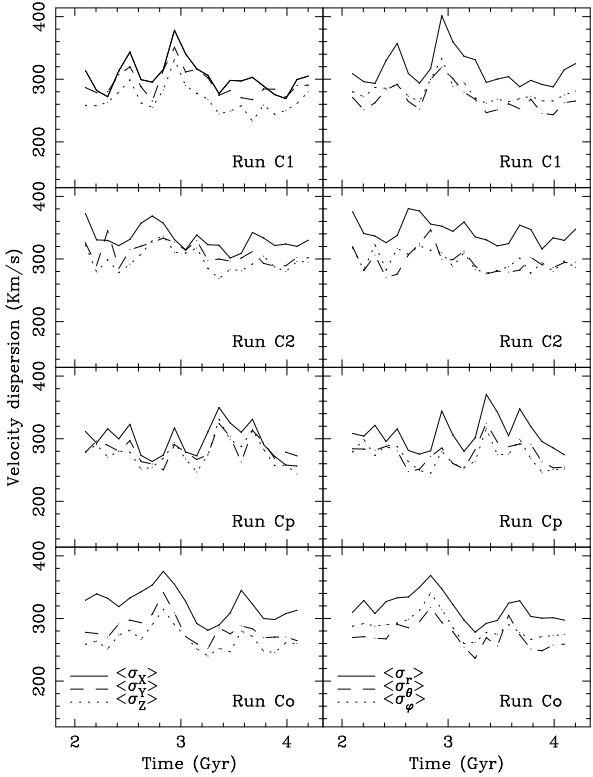}}

\caption{  Time evolution of the mean value of the velocity dispersion of
the central object formed in the collapsing simulations. In the left panel we 
show the time evolution along the three principal directions. 
These values indicate that the central galaxies have 
anisotropic velocity dispersion tensors. In the right
panel we show the median value of the dispersion using spherical coordinates.
Radial motions dominate in the objects formed by collapse.
}

\end{figure}
\begin{figure}

\epsfxsize=3.5truein
\centerline{\epsffile{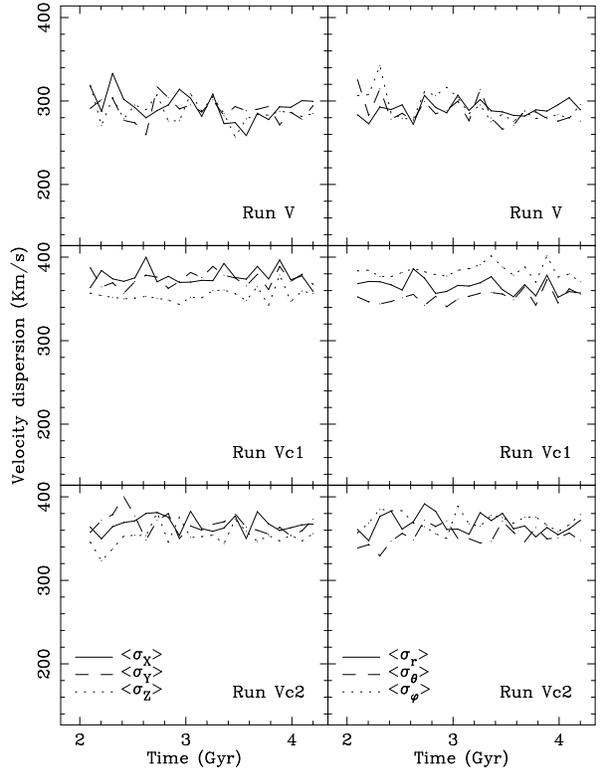}}

\caption{ Same as Fig.~14 but for the central objects formed in the
virialised simulations. In this case there are no differences between the
velocity dispersion along any of the principal directions, in agreement with
the fact that these objects are nearly spherical. Contrary to the case of
collapsing simulations, radial motions do not dominate, indicating a higher
degree of isotropy.
 }

\end{figure}

\subsection{Two dimensional properties.}

In order to compare our simulations with the observations of cD 
galaxies we use the following procedure.
We first choose a random projection of the central object.
We fix the $Z$ axis and make a rotation about it with a 
random angle between $0$ and $2\pi$. Then, we fix the $Y$ axis and repeat the 
same procedure and finally we do the same with the $X$ axis. Next, we select a 
random number between $0$ and $1$ and if this number is less than $1/3$ we 
project the galaxy on the $Y-Z$ plane, if the number is greater than $1/3$ and 
less than $2/3$ we project the galaxy onto the $X-Z$ plane and, if the 
random number is greater than $2/3$ we project onto the $X-Y$ plane. Then, for 
each particle, we keep its projected position and the vertical velocity. 
The projected object is placed at its center of mass according to these 
two dimensional positions and the two dimensional inertia tensor is calculated.
Using the two eigenvalues $(a>b)$ of this tensor we define for each particle 
the quantity 
$$ q = x^2 + \frac{y^2}{(b/a)^2}$$ 
\noindent
and the particles are ordered according to this value in increasing order. Then 
they are grouped in bins of $200$ particles and we compute the surface density
of each bin, except for the particles in the innermost $1.5$ kpc, which are the
ones mainly affected by the softening of our simulations. We do this for 
$9$ random projections of the central object in each simulation and for each 
timestep. This procedure allows us to study the 
time evolution of the projected density profiles of the central galaxies formed 
in our simulations, while checking at the same time for possible dependencies 
on the viewing angle. 

\subsubsection{Time evolution of the surface density profiles.}

The main bodies of D and cD galaxies 
have surface brightness profiles which are well fitted  by a de 
Vaucouleurs law (Lugger 1984, Schombert 1986). 
cD galaxies show an additional
luminous halo and the external parts of these galaxies no longer follow
the same de Vaucouleurs law as their main bodies (Oemler 1976, 
Schombert 1986). The colour profiles of these
halos seem to be essentially flat  and there is no evidence for breaks or 
discontinuities at the start of the cD envelope, nor 
for excessive blue colours in the envelope itself (Mackie 1992).

In order to be able to compare our results with the observations we study the time 
evolution of the surface density profiles using as a
reference the $r^{1/4}$ law.  We start our study at the time corresponding to
half the total time span of the simulation. At this time the central object has 
already formed and contains more than $10000$ particles.  
Using the procedure described above we compute the surface 
density profiles of the central galaxies formed in our simulations. These 
two-dimensional density profiles can be grouped into three categories. In the 
first category we find the profiles that can be well described by the 
de Vaucouleurs law.
These are the typical profiles of elliptical galaxies, but they are also typical
of the brightest cluster members found in poor AWM and MKW clusters and in some Abell 
clusters, for example NGC 2329 in A569, or the central galaxy in A2029 (see 
the Schombert 1986 profiles). 
In the second category we find the profiles that fall systematically below the 
$r^{1/4}$ law. 
This is also the case for some brightest cluster members, like the ones in 
A665, A1228 and  A2052 (Schombert 1986). Finally 
we come to the category of profiles typical of cD galaxies. In a galaxy's 
external parts, the profiles in this category are systematically above the 
$r^{1/4}$ law. This is the case for
the central galaxies in A779, A1413 and A2199 (Schombert 1986). 

We find that the density profile of the central galaxy is 
determined by the initial conditions of the simulation and does not depend on
the viewing angle. The tightly bound and virialised groups (runs Vc1 and Vc2) give 
central objects that can be well described by the $r^{1/4}$ law. The time 
evolution of the density profile of run Vc2 is shown in Fig.~16. 
The surface density profile of this galaxy is well fitted at all times 
by a de Vaucouleurs law and the same holds for the 
central galaxy formed in run Vc1. As shown in the previous section, the three 
dimensional profile of these objects 
is well described by the Hernquist profile. As the projection of the
Hernquist profile 
gives good fits to the $r^{1/4}$ law for a large range radii 
(Hernquist 1990) these good fits are not surprising. 
Note also the good agreement between the profiles obtained in the 
different projections at each timestep, indicating that there is no dependency 
of the surface density profile on the viewing angles.  
\begin{figure}

\epsfxsize=3.5truein
\centerline{\epsffile{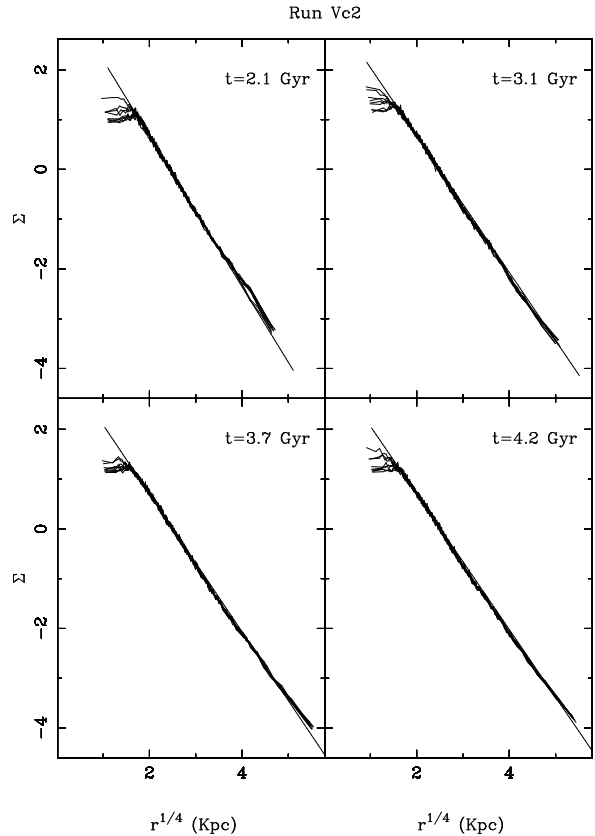}}

\caption{Time evolution of the surface density profiles of the central
object formed in run Vc2. For each time we show the profiles of nine random
projections and the mean best fitting de Vaucouleurs law, which is a good
description during all the times.
 }

\end{figure}

The second category of surface density profiles arises in the collapsing 
simulations, especially the ones with anisotropic initial conditions. 
Fig.~17 gives the time evolution of the profile of the central galaxy of run~Cp. 
We can see that the $r^{1/4}$ law gives good fits only in the main
parts of the galaxy, while the external parts fall systematically below this 
law. This is true for all projection angles, but is more pronounced for the 
profile along the minor axis and less so for the profile along the major axis.
The profiles of the central objects formed in the spherically collapsing 
simulations also have this feature, but it is not as pronounced, and the 
profiles can be well fitted by 
a $r^{1/4}$ at some timesteps. As the objects formed in the collapsing 
simulations are the ones which show more signs of triaxiality, especially in 
the anisotropic collapses and in the external parts, this behaviour
may be an effect of the triaxiality of the central galaxies. 
Thus, the 
presence of profiles falling below the $r^{1/4}$ law may be indicative of 
objects with strong departures from spherical symmetry.
\begin{figure}

\epsfxsize=3.5truein
\centerline{\epsffile{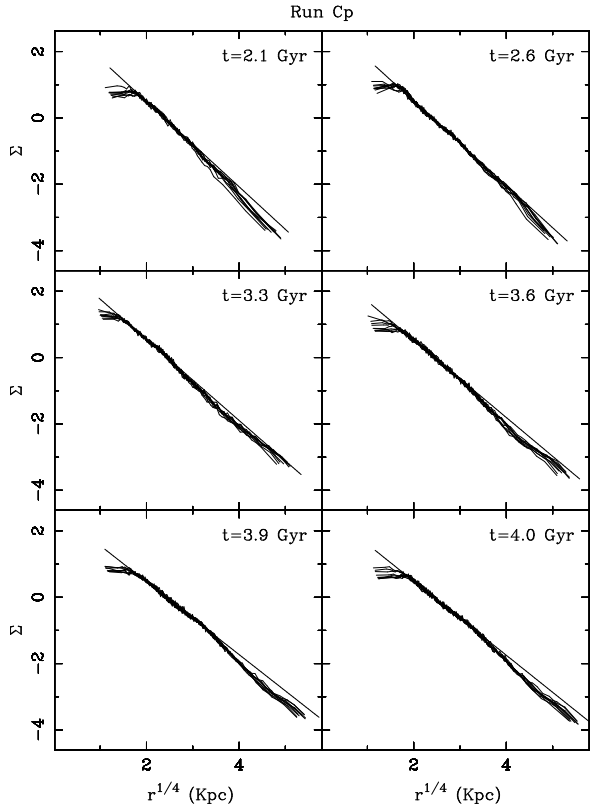}}

\caption{Same as Fig.~16 but for the object formed in run~Cp. In this case 
the $r^{1/4}$ law is a good fit only for the main body of the object,
while the external parts fall systematically below this law. This can be a
signature of the triaxiality of these objects.
 }

\end{figure}

The most interesting cases belong to the third type of surface density 
profiles, the ones typical of cD galaxies, shown in Fig.~18. 
These profiles are obtained only in the simulation of the more extended 
virialised group (run~V).
The central object formed in this simulation displays strong differences between
the outer shell of material and the inner parts. Moreover, as we saw in the
previous section, its three dimensional
density profile is not well described by the Hernquist law. This is a result of
the particular formation process of this object, where the mass coming from
stripped material is more important, and leads to surface density profiles 
typical of cD galaxies.
It is important to note that such profiles are not transient, as was the 
case in the simulations of merging galaxies by Navarro (1990), and that they 
are 
independent of the viewing angles. In our simulations, the deviation from a
single
$r^{1/4}$ law appears as the central object is formed. The inner parts, which 
correspond to the most bound particles, are well fitted by an $r^{1/4}$ law, 
while
the external parts, which correspond mainly to accreted material, form a halo 
that can be associated with the halos of cD galaxies found in the central parts
of clusters of galaxies. The mass of the system is distributed evenly 
between the central parts, well fitted by an $r^{1/4}$ law, and the external 
parts forming the halo.

\begin{figure}

\epsfxsize=3.5truein
\centerline{\epsffile{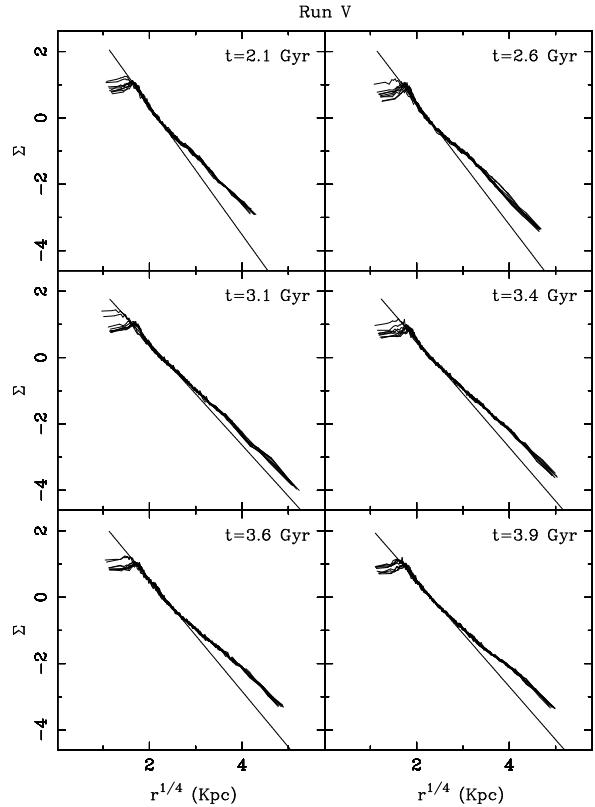}}
\caption{ Same as Fig.~16 but for the object formed in run V. In this
case the $r^{1/4}$ law fits the profile only in the main body of the object, 
while the
external parts have systematically higher values. This is the profile typical
of a cD galaxy. This effect is not a transient phenomenon and
is linked to the structure of the central galaxy.
 }

\end{figure}

\subsubsection{Position in the $\mu_e-R_e$ plane.}

One of the best studied correlations between the global
properties of elliptical galaxies is the relation between the parameters 
defining the best fitting $r^{1/4}$ law, i.e. the effective radius $R_e$ and 
the effective surface brightness $\mu_e$. These two parameters are found 
to be mutually dependent, with a relation of the form 
$\mu_e \simeq 3.3 \log R_e +$ constant (Kormendy 1977). Brightest cluster 
members seem to be an extension of the elliptical sequence towards greater 
effective radii and lower effective surface brightness. These galaxies,
however, have a tendency to be located above the mean relation defined by 
normal ellipticals (Schombert 1987) and even to have a shallower slope in this
relation (Hoessel et al. 1987). 

How are the central objects formed in our simulations 
distributed in the $\mu_e-R_e$ plane? In the preceding subsection we 
discussed the fits of the $r^{1/4}$ law to the projected density
of our central objects at different timesteps. From these we obtain 
the values of the 
corresponding $\mu_e$ and $R_e$,
assuming an $M/L = 5$. We used nine different random projections, 
but, since their results are very similar, we randomly select one 
of the set of values and we 
plot it on the $\mu_e -R_e$ plane. 
We note that different values of $M/L$ will of course shift the points
along the $Y$ axes, while maintaining their relative positions. This can 
also be 
achieved by another rescaling of the computer units. We will thus be mainly 
interested in the slope of the correlation.
The objects formed in our simulations follow a relation in this plane similar 
to the relation for elliptical galaxies. In the top panel of Fig.~19
we show the correlation for the central
galaxies formed in the collapsing groups and in the bottom panel the
correlation for the central galaxies formed in the virialised systems.
As the simulation evolves the central objects get denser and more extended and
the corresponding points in this $\mu_e-R_e$ plane are displaced towards 
greater $R_e$ and to smaller $\mu_e$. Both
groups of galaxies follow a relation of the form $\mu_e \simeq 3.7 
\log R_e +$ constant. This slope is somewhat higher than the one found for 
elliptical galaxies. Using all the data together we obtain a correlation with 
a slope of $3.9$. The objects formed under virialised initial conditions are 
less dense objects and so they fall systematically towards higher surface 
brightnesses than the objects formed in 
collapsing systems and this bias gives 
the higher slope when all the data are used together.
\begin{figure}

\epsfxsize=3.5truein
\centerline{\epsffile{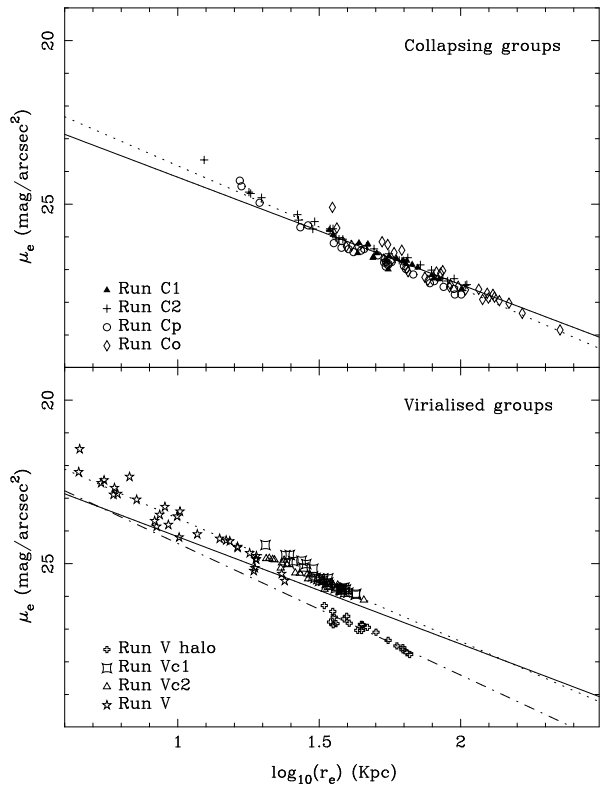}}

\caption{ Correlation between the surface brightness $\mu_e$ and 
effective radii $R_e$ for the central galaxies of our simulations at 
different timesteps. In the 
top panel we show the relation for the galaxies formed in collapsing groups 
and in the bottom panel the relation for galaxies formed under virialised 
initial conditions. The solid line is in both cases a line with the same slope 
as the Kormendy (1977) relation. The dotted line is the correlation for our
data. The dot-dashed line in the bottom panel shows the correlation for the 
halos of the
cD-like objects formed in run V. Symbols for collapsing groups: run~C1 filled triangles, run~C2 crosses, run~Cp circles, run~Co diamonds. Virialised groups: run~V halo swiss crosses, run~Vc1 lozenges, run~Vc2 triangles, run~V stars. 
 }

\end{figure}

Another interesting point in this respect concerns the halos of cD galaxies.
The properties of these halos can be characterised by fitting an $r^{1/4}$ law 
to the outer parts of the surface brightness profile, i.e. the part outside the 
region which is well fitted by the $r^{1/4}$ corresponding to the main body of
the galaxy. Schombert (1988) finds that, on the $\mu_e-R_e$ plane, these halos 
form an extension of the relation 
found for ellipticals and brightest cluster galaxies towards still lower
surface brightnesses and larger effective radii, perhaps with a steeper slope.
We repeated this for the halos of the central galaxies formed in our run V and
give the results in the bottom panel of Fig.~19. They have the same properties 
with respect to their parent objects as the halos of cD galaxies with 
respect to their parent galaxies. Schombert (1988) has argued that these halos, 
which also follow the luminosity profiles of other material in the cluster, 
like the diffuse background, can not form by mergers but have to form by a 
process separate from that of first-ranked 
ellipticals. This is not borne out by our simulations 
which show that, although the halo has in many respects different properties 
from the main body of the galaxy, there is no distinct discontinuity in the
formation process.
Schombert models these halos as a 
separate entity using a two-component model combining an elliptical galaxy and 
a separate halo component with different $M/L$ ratios and velocity dispersions, 
but the models are fitted with a wide range of values for these parameters. 
If the halo is a separate entity, we would expect it to have 
the same velocity dispersion as the system of secondary galaxies in the 
cluster. At first,
data from the central galaxy in A2029 (Dressler 1979) seemed to be in 
agreement with this idea. Recent data, however, suggest that, while the central
galaxy in A2029 does have a rising velocity dispersion profile, this is not a 
feature common to first-ranked galaxies (Fisher et al. 1995). The projected
velocity dispersion profiles of the galaxies obtained in our simulations are 
in agreement with the profiles of real brightest cluster members. This can be 
seen in Fig.~20, where we show the profile of the central galaxy formed in run~V
at the end of the simulation. This is a mean profile obtained by 
adding the 
profiles of the nine random projections of this object. The error bars 
indicate the dispersions over the mean values.  
The profiles for the rest of the galaxies obtained in our simulations are of 
the same nature and are independent of the viewing angles. 
The gradient in velocity dispersions is also in agreement with the gradients 
in the profiles of real galaxies. Thus, our simulations suggest that, 
the material that forms the halo of cD galaxies does not need to be
material with high velocity dispersion.
Deeper observations are needed to confirm this result.

\begin{figure}

\epsfxsize=3.5truein
\centerline{\epsffile{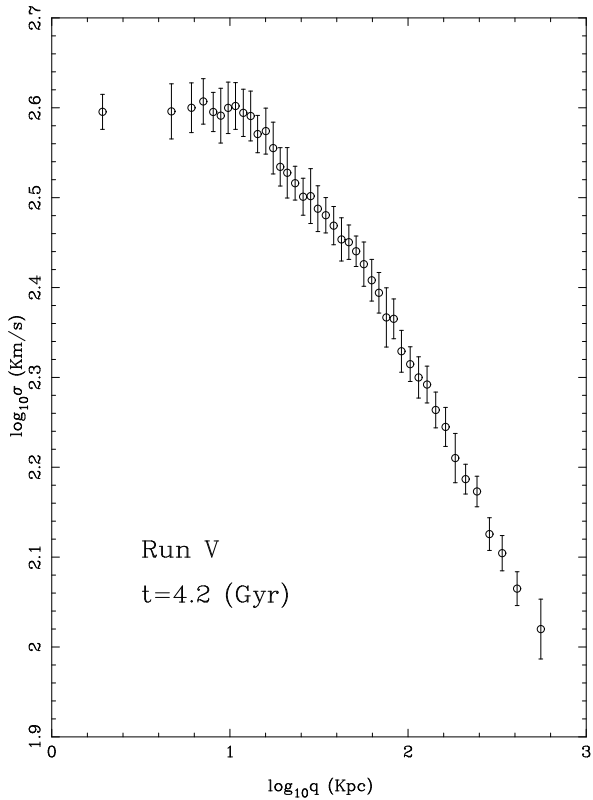}}

\caption{ Projected velocity dispersion profile of the central galaxy
formed in run V at the end of the simulation. The gradient in velocity
dispersion is comparable to the gradient found for real galaxies. The
profiles for the rest of the galaxies in our simulations are of similar
nature.
 }

\end{figure}

\subsubsection{The Faber-Jackson relation.}

The Faber-Jackson relation (Faber \& Jackson 1976) is a relation of the 
form $L \sim \sigma^p$ between the total luminosity of elliptical
galaxies $L$, and their central velocity dispersions $\sigma$.
The value of $p$ is still controversial but the most commonly accepted
one is $p=4 \pm 0.7$ (Terlevich et al. 1981). The brightest cluster members
do not follow this correlation very well, and tend to be brighter
than predicted from their central velocity dispersions using the relation
$L\sim \sigma^4$ (Malumuth \& Kirshner 1981, 1985).

The relations for the galaxies formed in our simulations are shown in
Fig.~21. Instead of luminosity we use the total mass. This seems to be a good 
approximation, as the $M/L$ ratio for ellipticals seems to be independent of 
luminosity (Tonry \& Davis 1981), or a weakly dependent function of the 
luminosity of the form $M/L \sim L^{1/4}$ (Oegerle \& Hoessel 1991). 
Objects formed in collapsing simulations are located on the 
$\log_{10}M-\log_{10}\sigma$ plane very differently from 
the objects formed under virialised initial conditions. 
The galaxies formed in collapsing groups 
do not follow a Faber-Jackson relation and give a scatter diagram in the 
$\log_{10}M-\log_{10}\sigma$ plane, 
while the data corresponding to the galaxies
formed from virialised initial conditions show much less scatter. This can be
explained if the Fundamental Plane is a consequence of the virial
theorem (Pahre et al. 1995). As we have seen in Fig.~14, the velocity 
dispersion profiles of these galaxies indicate that these systems are not
in virial equilibrium. On the other hand, the galaxies
formed under virialised conditions are fully isotropic systems and give
better correlations. The solid line shown in both diagrams corresponds to a 
line with the same slope as the Faber-Jackson relation. The dashed line
shown in the panel of collapsing groups is a least squares fit, while the
dashed line in the 
panel of virialised groups corresponds to the least squares fit of the
galaxies formed in runs Vc1 and Vc2. These objects, which 
are fully virialised systems, give a slope of $3.6$, i.e. in the range of 
the Faber-Jackson relation. This value, however, is very uncertain, as can 
be seen from the location of the corresponding points in the lower panel of 
Fig. 21.  
It is interesting to note that the objects formed in run~V, which can be 
associated with
the cD galaxies in clusters, fall systematically above the line corresponding
to the correlation for runs Vc1 and Vc2 which resemble elliptical galaxies, as
is the case for real cD galaxies (Schombert 1987). As stated in the beginning of this section, 
Malumuth \& Kirschner (1985) find that brightest cluster members are 
systematically brighter than what could be expected by the Faber-Jackson 
relationship. They furthermore find that this effect is stronger for the 
subset of their galaxies classified by Morgan and his coworkers as cD. It is 
tempting to draw an analogy between this result and our simulations. 
Unfortunately the remainder of the Malumuth \& Kirschner sample could also 
contain some cD galaxies. Thus more observational work is needed for a better 
comparison.
\begin{figure}

\epsfxsize=3.5truein
\centerline{\epsffile{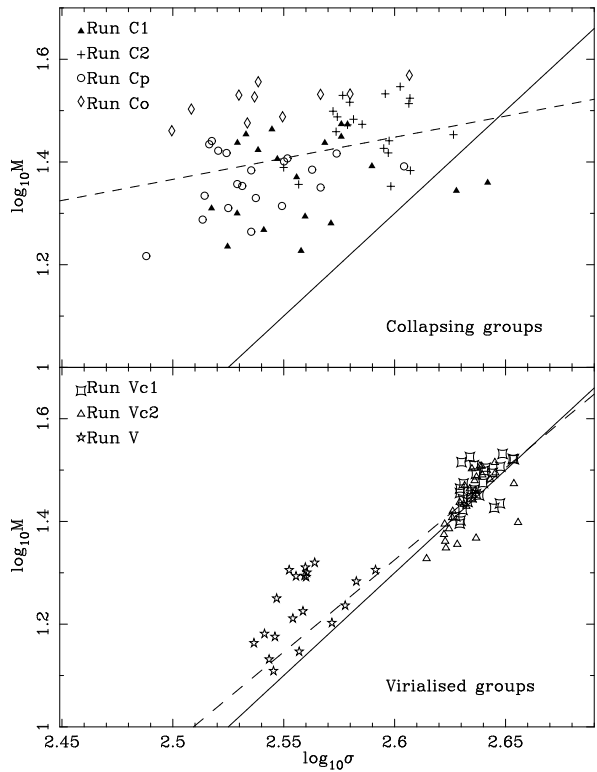}}

\caption{ Faber-Jackson relation for the central galaxies in our
simulations. In the top panel we show the results for the galaxies formed in
the collapsing groups and in the bottom panel the results for the galaxies
formed in virialised groups. The solid line is a line with a similar slope as
the one for elliptical galaxies. The dotted line is the correlation obtained
from our data. The different symbols correspond to different simulations
and several timesteps are shown for each simulation. Symbols as in Fig.~19.
 }

\end{figure}

\section{Summary.}

We have performed N-body simulations of the dynamical evolution of groups of 
galaxies with a variety of initial conditions and with all the mass initially 
bound to galaxies. Simulations with a common halo encompassing the group 
will be discussed in a forthcoming article. 
Some of the simulations correspond to free collapses with and without spherical
symmetry and the rest to initially virialised systems. Our results are 
relevant to large groups or 
poor clusters, subcondensations within larger clusters, during times when 
the influence of the surrounding cluster can be neglected, or to subunits 
that have come together during cluster formation time to form a large cluster.

The dynamics of all but one of these systems is driven by the
merging instability (Carnevali et al. 1981). We find that the
condition for this instability to be operative in virialised systems is that 
there must be a central concentration of matter that drives the orbits of the
galaxies to the center of the system. When this condition is fulfilled, both
collapsing and virialised groups form a giant central object in the center.
In this paper we have studied its properties as a function of the initial conditions 
of the simulations.

These central objects grow in time using two different mechanisms. The first
is known as ``galactic cannibalism'' and consists in the merging of galaxy
satellites that pass near the giant galaxies. The second one
is the accretion of galactic material that is stripped from the satellite
galaxies by tidal forces. The prevalence of one or 
another mechanism depends on the initial conditions of the group. 
In the more extended and virialised systems the rate of mass 
increase due to tidal stripping is comparable to the mass growth due to
merging. In the rest of the simulations that form a central galaxy, the mass 
growth by merging is more important.

The objects formed are oblate or mildly triaxial in nature, especially
in the cases of collapsing
groups with aspherical initial conditions. In such cases the orientation 
of the central object correlates well with that of the initial group. In 
general the triaxiality is stronger in the outer parts of the central 
object, in good agreement with observations (Porter, Schneider \& Hoessel 
1991; Mackie, Visvanathan \& Carter 1990). Collapsing systems are supported 
by anisotropic velocity dispersion
tensors. The tightly bound and virialised groups form nearly spherical central 
galaxies. In most cases these systems can be 
described as a stratification of ellipsoids with the same axial ratios. These 
galaxies have isotropic velocity dispersion tensors. In these cases the 
volumetric density is  well fitted by an ellipsoidal Hernquist profile.

Projecting the density distribution, in order to compare them with the properties of ellipticals and brightest cluster members, we obtain surface density profiles that
can be compared with the surface brightness profiles of real galaxies. We
obtain three types of profiles that are representative of the profiles of the 
galaxies in the centers of clusters. Tightly bound virialised clusters
and spherical collapses give objects with profiles well fitted by an $r^{1/4}$
law. These results are in agreement with the simulations of van
Albada (1982) and May and van Albada (1984). The collapsing simulations
from aspherical initial
conditions give galaxies whose surface density profiles are well fitted by an 
$r^{1/4}$ law only in the main body of the object. The surface density in the 
external parts falls below the $r^{1/4}$ law. This kind of
profile is also observed in real brightest cluster members. As the galaxies
formed in these simulations are non-spherical objects, we suggest that
this could be
a signature of the triaxiality in the surface density profiles. The most
interesting case corresponds to the central galaxy formed in the extended
virialised clusters, where stripping is more important. The surface density 
profile of this object is also well fitted by an $r^{1/4}$ law 
in the main body of the galaxy, but in the external parts the surface density
profile is systematically above this
reference law, as is the case for cD galaxies in clusters. 
The central objects in our simulations have projected velocity dispersion 
profiles that are
comparable to the profiles of real galaxies. They seem to have a greater
luminosity
than the luminosity corresponding to the central velocity dispersions given by
the Faber-Jackson relation and thus again reproduce a property of brighter
cluster members. 

\begin{acknowledgements}

We thank A. Bosma and the referee, A. Romeo, for their helpful
comments.  We also wish to thank Lars Hernquist for kindly providing
us with the vectorised version of the treecode, and the ``Institut de 
d\'eveloppement et des resources en 
informatique scientifique, Orsay, France (I.D.R.I.S.)" for the allocation of 
computer time. Part of this
work was supported by the {\it Direcci\'on General de Investigaci\'on y
Tecnolog\'{\i}a} under contract PB 93-0824-C02-02.

\end{acknowledgements}


\begin{thebibliography}{}

\bibitem{} Andreon, S. , Garilli, B., Maccagni, D., Gregorini, L., Vettolani, G., 1992, 
A\&A 266,127
\bibitem{} Austin, T.B., Peach, J.V., 1974, MNRAS 168, 591
\bibitem{} Barnes, J., 1992, ApJ 393, 484
\bibitem{} Barnes, J., Hernquist, L., 1992, ARA\&A 30, 705
\bibitem{} Barnes, J., Hut, P., 1986, {\it Nature} 324,446
\bibitem{} Beers, T.C., Geller, M.J., 1983, ApJ 274, 491
\bibitem{} Binggeli, B., 1982, A\&A 107, 338
\bibitem{} Bird, Ch. M., 1994, AJ 107, 1637
\bibitem{} Blakslee, J.P., Tonry, J.L., 1992, AJ 103,1457
\bibitem{} Bode, P.W., Berrington, R.C., Cohn, H.N., Lugger, Ph. M., 1994, ApJ 433, 479
\bibitem{} Carnevali, P, Cavaliere, A., Santangelo, P., 1981, ApJ 249, 449
\bibitem{} Carter, D., Inglis, I., Ellis, R.S., Efstathiou, G., Godwin, J.G., 1985, MNRAS 
212,471
\bibitem{} Carter, D., Metcalfe, N., 1980, MNRAS 191, 325
\bibitem{} Cowie, L.L., Binney J., 1977, ApJ 215, 723
\bibitem{} Dressler, A., 1979, ApJ 231, 659
\bibitem{} Dubinski, J., Carlberg, R.G., 1991, ApJ 378, 496
\bibitem{} Faber, S.M., Jackson, R.E., 1976, ApJ 204, 688
\bibitem{} Fabian, A.C., Nulsen, P.E.J., 1977, MNRAS 180, 479
\bibitem{} Fisher, D., Illingworth, G., Franx, M., 1995, ApJ 438,539 
\bibitem{} Funato, Y., Makino, J., Ebisuzaki, T., 1993, PASJ 45, 289
\bibitem{} Gallagher, J.S., Ostriker, J.P., 1972, AJ 77,288
\bibitem{} Garc\'{\i}a-G\'omez, C., Athanassoula, E., Garijo, A., 1996, A\&A 313, 363
\bibitem{} Hernquist, L., 1988, ApJS 64,715
\bibitem{} Hernquist, L., 1990, ApJ 356, 359
\bibitem{} Hill, J.M., Oegerle, W.R., 1993, AJ 106, 831
\bibitem{} Hoessel, J.G., Schneider, D.P., 1985, AJ 90,1648
\bibitem{} Hoessel, J.G., Oegerle, W.R., Schneider, D.P., 1987, AJ 94, 1111
\bibitem{} Kormendy, J., 1977, ApJ 218, 333
\bibitem{} Kormendy, J., Djorgovski, S., 1989, ARA\&A 27, 235
\bibitem{} Lambas, D.G., Groth, E.J., Peebles, P.J.E., 1988, AJ 95, 996
\bibitem{} Lauer, T.R., 1988, ApJ 325, 49
\bibitem{} Lugger, Ph. M., 1984, ApJ 286, 106
\bibitem{} Mackie, G., 1992, ApJ 400,65
\bibitem{} Mackie, G., Visvanathan, N., Carter, D., 1990, ApJ 73, 637
\bibitem{} Malumuth, E.M., Kirshner, R.P., 1981, ApJ 251, 508
\bibitem{} Malumuth, E.M., Kirshner, R.P., 1985, ApJ 291, 8
\bibitem{} Malumuth, E.M., Kriss, G.A., Dixon, W.V.D., Ferguson, H.C., Ritchie, C., 1992, 
AJ 104,495
\bibitem{} Malumuth, E.M., Richstone, D.O., 1984, ApJ 276, 413
\bibitem{} McNamara, B.R., O'Connell, R.W., 1992, ApJ 393,579
\bibitem{} May, A., van Albada, T.S., 1984, MNRAS 209, 15
\bibitem{} Merrifield, M.R., Kent, S.M., 1991, AJ 101, 783
\bibitem{} Merritt, D., 1983, ApJ 264, 24
\bibitem{} Merritt, D., 1984, ApJ 276, 26
\bibitem{} Merritt, D., 1985, ApJ 289, 18
\bibitem{} Morbey, Ch., Morris, S., 1983, ApJ 274, 502
\bibitem{} Mould, J.R., Oke, J.B., de Zeeuw, P.T., Nemec, J.M., 1990, AJ 99, 1823
\bibitem{} Navarro, J.F., 1990, MNRAS 242, 311
\bibitem{} Oegerle, W.R., Hill, J.M., 1992, AJ 104, 2078
\bibitem{} Oegerle, W.R., Hoessel, J.G., 1991, ApJ 375, 15
\bibitem{} Oemler, A., 1976, ApJ 209, 693
\bibitem{} Ostriker, J.P., Hausman, M.A., 1977, ApJ 217,L125
\bibitem{} Ostriker, J.P., Tremaine, S.D., 1975, ApJ 202,L113
\bibitem{} Pahre, M.A., Djorgovski,S.G., de Carvalho, R.R., 1995, ApJ 453, L17
\bibitem{} Porter, A.C., Schneider, D.P., Hoessel, J.G., 1991, AJ 101, 1561
\bibitem{} Quintana, H., Lawrie, D.G., 1982, AJ 87,1
\bibitem{} Rhee, G., Katgert, P., 1987, A\&A 183, 217
\bibitem{} Rhee, G., Roos, N., 1990, MNRAS 243, 629
\bibitem{} Richstone, D.O., 1975, ApJ 200, 535
\bibitem{} Richstone, D.O., 1976, ApJ 204, 642
\bibitem{} Richstone, D.O., Malumuth, E.M., 1983, ApJ 268, 30
\bibitem{} Romanishin, W., 1987, ApJ 323, L113
\bibitem{} Rood, H.J., Sastry, G.N., 1972, AJ 77, 451
\bibitem{} Ryden, B.S., Lauer, T.R., Postman, M., 1993, ApJ 410, 515
\bibitem{} Sastry, G.N., 1968, PASP 80, 252
\bibitem{} Schombert, J.M., 1986, ApJS 60, 603
\bibitem{} Schombert, J.M., 1987, ApJS 64, 643
\bibitem{} Schombert, J.M., 1988, ApJ 328, 475
\bibitem{} Schombert, J.M., 1992, in ``Morphological and Physical Classification of 
Galaxies". G. Longo et al. eds. Kluwer Academic Publishers, p. 53-68
\bibitem{} Sharples, R., Ellis, R., Gray, P.M., 1988, MNRAS 231,479
\bibitem{} Struble, M.F., Peebles, P.J.E., 1985, AJ 90, 582
\bibitem{} Terlevich, R., Davies, R.L., Faber, S.M., Burstein, D., 1981, MNRAS 196,
381
\bibitem{} Thuan, T. X., Romanishin, W., 1981, ApJ 248, 439
\bibitem{} Tonry, J.L., 1987, in ``Structure and Dynamics of Elliptical Galaxies''. T. 
de Zeeuw (ed.) Proc. IAU Symp. 127, Reidel, Dordrecht, p. 89-98
\bibitem{} Tonry, J., Davis, M., 1981, ApJ 246, 680
\bibitem{} van Albada, T.S., 1982, MNRAS 201, 939
\bibitem{} Zabludoff, A.I., Huchra, J.P., Geller, M.J., 1990, ApJS 74, 1

\end{thebibliography}
\end{document}